# Fano Resonances of Plasmonic Nanodisk

*Zheyu Fang[1], Junyi Cai[1], Zhongbo Yan[1], Nathaniel K. Grady[4], Xing Zhu[1,2]*

[1]School of Physics, State Key Laboratory for Mesoscopic Physics, Peking University, Beijing 100871, China

[2]National Center for Nanoscience and Technology, Beijing 100190, China

[3]Beijing National Laboratory for Condensed Matter Physics and Institute of Physics, Chinese Academy of Science, Box 603-146, Beijing 100190, China

The Fano resonance of a single symmetry broken Ag nanodisk under a normal incidence was investigated by using finite-difference time-domain (FDTD) simulations. The asymmetry line shape of the Fano resonance was controlled by modifying the open angle of the nanodisk, and this Fano splitting was demonstrated as the result of the overlap between the broad dipolar and narrow quadrupolar modes, which could be strengthened by enlarging the radius of the nanodisk. A semi-analytical method was developed to calculate the plasmon hybridization, which was used to analyze the sub-process of the quadru Fano resonance. With the good agreement between theoretical calculations and FDTD simulations, the suggested method provides a way to investigate and control the Fano resonance inside a single planar nanostructure, and can be applied to the future high-performance Fano resonance sensors.

Surface plasmons, as collective electrons oscillations at the interface between the metal and dielectric layers, are interest for tremendous emerging applications,[1,2] like subwavelength optical imaging and focusing,[3-5] plasmonic waveguiding,[6-10] and a broad range of chemical and biomolecular sensing.[11,12] For the recent research of plasmonics, coherent phenomena related to the Fano resonance,[13-15] supper/sub-radiance,[16-18] and plasmon induced transparency[19-22] have become hot topics for the future applications, such as the resonance line shape control.[23] The hybridization of elementary plasmons supported by nanostructures of elementary geometries was described by an electromagnetic analog of molecular orbital theory.[24] The metallic nanosphere,[25] nanorice,[26] and nano-egg[27] were investigated as reduced-symmetry nanostructures for the splitting of plasmon modes and the onset of electroinductive plasmons upon controlled. The core-shell nanostructures[28-30] and dimer plasmons[31,32] were attracted a dramatic attention for the effort of the plasmon resonance line shape control and the application for the molecule biosensing.

The bonding and antibonding plasmon modes are generated by the interaction of parent plasmon modes of individual nanostructures. The concentric ring/disk cavity (CRDC) configuration[30] was proposed and investigated for their plasmon narrow dark modes (subradiance) and broad bright modes (superradiance) hybridized by the dipole of the ring and disk cavities, respectively. This kind of plasmon hybridization was further modified by introducing the idea of symmetry breaking as the offset between centers of ring and disk cavities known as the nonconcentric ring/disc cavity (NCRDC) structure.[33] Fano resonances arise directly from the coherent interference of bright and dark plasmon modes. Quadrupolar (or octupolar) Fano resonance could also be generated when the quadrupole (or octupole) of the ring cavity overlaped with the broad bright mode. Besides the NCRDC, nanoparticle clusters,[14, 34] such as the heptamer structure, also exhibited strong magnetic and Fano-like resonances by tailoring the number and position of the sphere in close-packed clusters. The physics insight of Fano resonances is the plasmon coherent interference, and it has been demostrated to be significantly sensitive to the reducing degree of the structure symmetry. The plasmon hybridization requires parent plasmon modes, supported by individual nanostructures, to generate bright and dark modes. And the coherent interference needs the modification of the narrow dark mode to superposition with the bright mode so as to generate a Fano resonance.

For a metallic nanodisk in vacuum, there was no coupling between the bright dipole and dark quadrupole. However, when the nanodisk was placed on a dielectric substrate,[35, 36] the image charge of the plasmon formed a significant quadrupolar field across the nanostructure, introducing the hybridization between its dipolar and quadrupolar plasmons, and resulted a quadrupolar Fano resonance. Nanoshells were also reported to be deposited on high dielectric permittivity substrates to induce a weak Fano resonance.[35] And the sensitivity of localized surface plasmon resonance (LSPR) was further improved when a plasmonic nanocube was deposited on a dielectric substrate.[36]

In this letter, we report the line shape control of the Fano resonance inside a single metallic nanodisk by reducing its structure circular symmetry. We analyze the plasmon extinction of this symmetry broken nanodisk with different open angles, disk radii, and incident laser orientations. We further calculate the plasmon hybridization between the generated dark and bright dipoles by using a semi-analytical method with finite-difference time-domain (FDTD) simulations. We demonstrate that the observed Fano resonance is induced by the superposition between a dark plasmon quadrupole and a bright plasmon dipole. Understanding the underlying physics of this picture opens a new path to design and control the Fano resonance of a single plasmonic nanoparticle.

The theoretical model of an Ag nanodisk with an open angle was performed using a commercial software (Lumerical Solutions) implementation of the FDTD method. The experimentally measured Ag permittivity data (CRC) was used in our simulations, and the structure was modeled in the frequency domain with an incident excitation as the total-field scattered-field source. In the simulation, we built two analysis groups, each of which consisted of a box power monitor outside the nanodisk, with one in the total field region and the other in the scattered field region, and the absorption and scattering cross sections were calculated by integrating the power flux over the enclosed box surface, respectively. The extinction then was obtained as the sum of the absorption and scattering data.

In Figure 1, extinction spectra of Ag unbroken and broken nanodisks are compared. The normal incidence with the polarization parallel to the planar nanodisk is used to excite plasmons. The model schematic is shown as the inset of the simulated extinction spectrum. The thickness for both nanodisks is 30 nm, and the open angle for the symmetry broken nanodisk is 90 degree. From the extinction curves, we can see that one of the hybridized plasmon modes is split into two peaks denoted as modes (*i*) and (*iii*). For the normal incidence, most of the induced surface charges will concentrate at the structure edge. The edge charge density at the right side of Figure 1 shows that both the mode (*i*) and (*iii*) are almost the same dipolar plasmons with the positive and negative charges at opposite sides. For the charge density map, the red and blue color presents the positive and negative surface charge, respectively. If we look at the charge plot of the trough (*ii*) between the peak (*i*) and (*iii*), it exhibits a different charge distribution, where the charge phase is turned over and the charge density shows a quadrupolar resonance. Besides, the line shape of mode (*iii*) is highly asymmetric with a steeper slope toward the red than toward the blue. And we find the asymmetric character of mode (*iii*) can also be well fitted with the Fano line shape (See Supporting Information). Thus, we consider that a narrow quadrupolar mode overlaps with the hybridized plasmon dipolar mode around 408 nm, turning over the phase of the edge charge at that point, and induces a quadru Fano resonance. On the other side, the edge charge of peak (*iv*) is the same as the one of the unique peak for the unbroken nanodisk, but with the peak position red shifted to the near inferred.

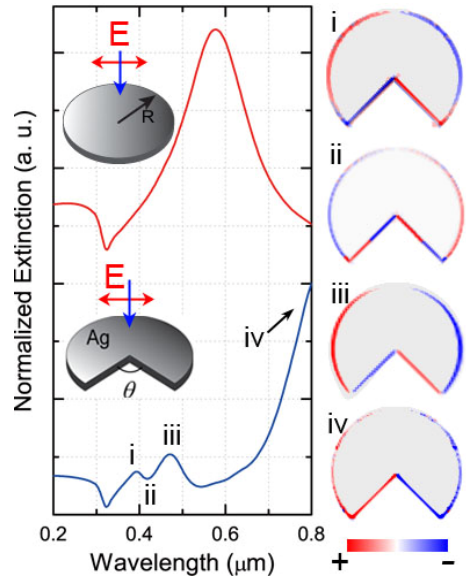

**Figure 1.** Normalized extinction spectra of Ag unbroken and broken nanodisks under an normal incident excitation calculated by FDTD simulations, the radius and thickness for both nanodisks are 80 nm and 30 nm respectively, the open angle is 90 degree; (i-iv) edge charge distributions for modes *i-iv* denoted on the extinction curve, calculated as the divergence of the simulated electric field, the red and blue color represents the positive and negative charge respectively.

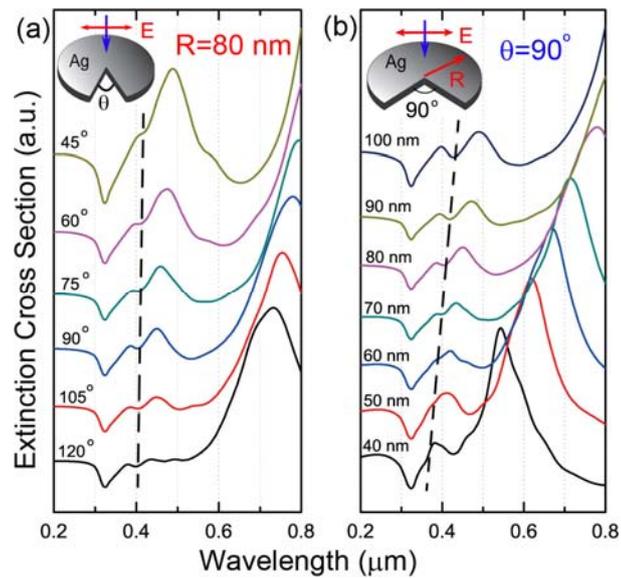

**Figure 2.** (a) Normalized extinction cross sections for an Ag nanodisk ($R$ = 80 nm) with different

open angles $\theta$ from 45 to 120 degree; (b) extinction spectra of the Ag symmetry broken disk ($\theta$ = 90 degree) with different disk riduii from 40 nm to 100 nm.

In order to investigate this asymmetric Fano resonance line shape, symmetry broken nanodisks with different open angles ($\theta$) and nanodisk radii ($R$) are investigated with their extinction spectra obtained by using FDTD simulations, as shown in Figure 2. The thickness of the nanodisk is kept as 30 nm unchanged. In Figure 2a, the radius of the nanodisk is modeled as 80 nm. By increasing of the open angle, the structure symmetry can be reduced, and the overlap between the narrow quadrupole and broad hybridized dipole can be controlled, see the dashed line in Figure 2a. For the open angle $\theta$ = 45 degree, the superposition between the quadrupole and dipole is not obvious, and only a small step appears on the simulated extinction spectrum. With $\theta$ increasing, the dark quadrupolar mode is enhanced by the increase of the equivalent charge interaction length between the two arms of the open angle, which magnifies the Fano interference and makes the asymmetry resonance line shape more obvious as $\theta$ = 90 and 105 degree. But with $\theta$ > 120 degree, though the equivalent charge interaction length is further enlarged, the quadrupole is retrenched because the reducing of the edge charge density. The normal incidence with a polarization perpendicular to the half separate line of the open angle will decrease the charge density as the open angle increasing (see the inset of the Figure 2a). Thus, there is a trade off between the open angle degree and the edge charge density, and the optimal Fano resonance can be obtained when $\theta$ = 90 degree.

To investigate how this Fano splitting happens, the radius ($R$) of the nanodisk is regulated from 40 to 100 nm. Figure 2b shows the simulated extinction spectra of the nanodisk with the open angle $\theta$ = 90 degree. For a smaller nanodisk ($R$ = 40 nm), because its quadruple is week, only a protrusion is observed around 360 nm on the extinction curve. With the increasing of the radius, this protrusion experiences a spectroscopic red-shift and presents a Fano splitting at 400 nm when the radius enlarged to 70 nm. The Fano splitting can be further enhanced when the nanodisk radius expends to 100 nm (see the dashed line in Figure 2b), where a strong overlap between the narrow quadrupolar and the hybridized dipolar modes takes place, and results an asymmetry Fano line shape with a steeper slope toward the red than toward the blue.

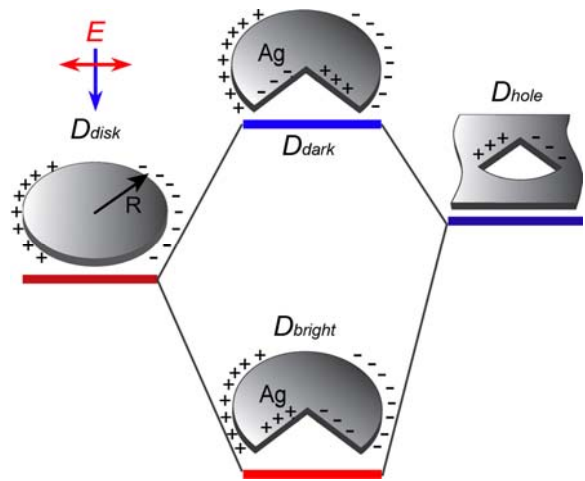

**Figure 3.** Schematic of the plasmonic hybridization of the dipolar modes generated by the disk and hole structure, and the energy diagram of the degenerated bright and dark modes.

To understand the physics insight of this Fano splitting, we start to analyze this single structure plasmon resonance with the hybridization between two individual nanostructures as an

Ag unbroken nanodisk (disk), and an infinite Ag layer with an inside fan hole (hole), as shown in Figure 3. The hybridization of elementary plasmons supported by disk and hole can be described by the electromagnetic analog of molecular orbital theory.[24] The primitive plasmon modes of the disk ($D_{disk}$) and hole ($D_{hole}$) in vacuum are calculated with FDTD simulations. Figure 3 shows the plasmon hybridization between the $D_{disk}$ and $D_{hole}$, and their generated new modes as $D_{bright}$ and $D_{dark}$, which have exactly the same edge charge distribution of the dipolar mode (*i*) and (*iv*) as shown in Figure 1. The energy diagram in Figure 3 also gives qualitatively how the disk-hole interaction results in the degenerate states as $D_{bright}$ and $D_{dark}$. The Fano splitting then can be considered as the quadrupole of the hole structure ($Q_{hole}$) overlapping with the $D_{dark}$ mode (compared to the quadrupole, the $D_{dark}$ is more superradiance), and generate a quadru Fano resonance.

In order to demonstrate this assumption, we develop a semi-analytical method with the help of FDTD simulations to calculate the plasmon hybridization between the $D_{disk}$ and $D_{hole}$ modes, which is a sub-process that cannot be obtained directly by either analytical formula calculations or FDTD numerical simulations. The model for the semi-analytical calculations is shown in Figure 4, where the normal incident laser with the polarization perpendicular to the unit vector *n* pointed from the center of the disk to the center of the hole, as shown in the right middle side of Figure 4.

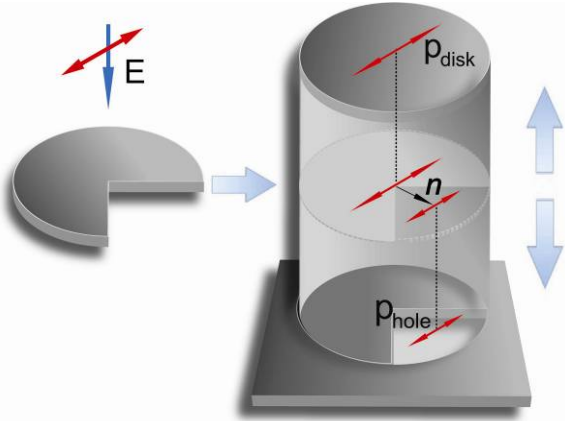

**Figure 4.** Schematic model of the semi-analytical calculation for the plasmon hybridization process. The unite vector *n* related to the center of disk and hole is perpendicular to the incident polarization.

As we known, the dipole can be expressed as $p=\alpha E$, where $\alpha$ is the electronic polarizability, and $E$ is the incident electric field. We define a new variable $Z = 1/\alpha$ to describe $\alpha$ in a dipole, thus $E = Zp$. If considering the total electric field on the disk ($Z_{disk} p_{disk}$) is the sum of the incident electric field ($E_{incident}$) and the interaction electric field induced by the hole structure ($Z_{inter} p_{hole}$), then we can get following equations

$$E_{incident} + Z_{inter} p_{hole} = Z_{disk} p_{disk}$$
$$E_{incident} + Z_{inter} p_{disk} = Z_{hole} p_{hole} \quad (1)$$

where $p_{disk}$ and $p_{hole}$ represent the dipole of disk and hole structure, respectively.

To calculate $Z_{disk}$ ($Z_{hole}$), we first consider the forward scattering electric field ($E_{forward}$) of the dipole in an individual disk (hole) structure. $E_{forward}$ then can be written as

$$E_{forward} = k^2 \frac{e^{ikr_1}}{r_1} p_{disk} = k^2 \frac{e^{ikr_1}}{r_1} \frac{E_{incident}}{Z_{disk}} \quad (2)$$

where $k$ is the incident wave vector, and $r_1$ is the distance between the simulated disk (hole) and the detected point, which in our calculation is 1.0 m. With FDTD simulations, the $E_{incident}$, $E_{forward}$, and $k$ can be obtained, and the dipolar coefficient $Z_{disk}$ ($Z_{hole}$) can be calculated.

To evaluate the interaction coefficient $Z_{inter}$, the electric field at the near-field of the dipole of the disk can be expressed as

$$E_{disk} = \frac{1}{r_2^3}[3n(p_{disk} \cdot n) - p_{disk}] \qquad (3)$$

Where $n$ is the unit vector pointed from the center of the disk to the center of the hole, as shown in Figure 4. In our calculation, $r_2$ represents the equivalent length between these two centers, and is supposed as $r_2=2/3R$, where $R$ is the radius of the disk and hole. Because the incident polarization is perpendicular to $n$ as denoted in Figure 4, the $E_{disk}$ then can be reduced as $E_{disk} = -p_{disk}/r_2^3$. From the equation $E=Zp$, we can obtain the interaction coefficient $Z_{inter}$ as $-1/r_2^3$.

Then the plasmon hybridization between $p_{disk}$ and $p_{hole}$ can be written as

$$p_{disk} + p_{hole} = \frac{Z_{disk} - Z_{hole}}{Z_{disk}Z_{hole} - Z_{inter}^2} E_{incident} \qquad (4)$$

By substituting the calculated dipolar coefficients $Z_{disk}$, $Z_{hole}$ and $Z_{inter}$, we can numerically plot the plasmon hybridization between the disk and hole dipolar modes, and their degenerate states as bright and dark modes, which is a sub-process before the quadru Fano resonance taking place. The semi-analytical calculated result is plotted in Figure 5b, where the resonance modes $D_{bright}$ and $D_{dark}$ are denoted on the curve.

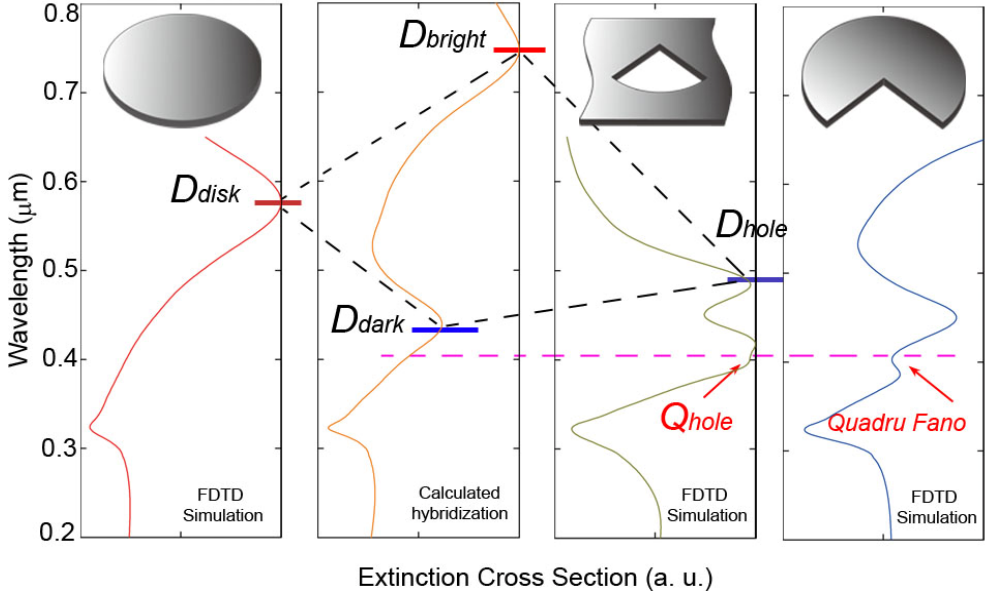

**Figure 5**(a) Extinction spectrum of the Ag disk, with its resonance peak ($D_{disk}$) locating at 579 nm; (b)calculated spectrum of the plasmon hybridization between $p_{disk}$ and $p_{hole}$; (c) extinction spectrum of the hole structure with a dipole ($D_{hole}$) and a quadrupole ($Q_{hole}$); (d) extinction spectrum for the symmetry broken disk with a Fano splitting at 408 nm. The thickness and radius of the tested sample are 30 and 80 nm, and the open angle is 90 degree.

In Figure 5, we show the plasmon hybridization diagram for the quadrupolar Fano resonance. The radius and thickness for all the tested samples are 80 nm and 30 nm, respectively. The open angle is chosen as 90 degree. The extinction spectrum of Figure 5a shows the dipolar mode of the disk with its resonance peak ($D_{disk}$) locating at 579 nm. The dipole ($D_{hole}$) and quadrupole ($Q_{hole}$) modes of the hole are also denoted in the extinction curve obtained by FDTD simulations as Figure 5c. As the energy diagram in Figure 3, the resonance dipole mode $D_{disk}$ and $D_{hole}$ can be hybridized and degenerate new modes as $D_{bright}$ and $D_{dark}$. They experience spectroscopic blue-shift and red-shift, respectively, which is confirmed by our semi-analytical calculation results as Figure 5b. The resonant peak for the calculated $D_{dark}$ mode in Figure 5b locates at 420 nm, and its full width at half maximum (FWHM) is about 80 nm. Because the peak of the quadrupolar mode ($Q_{hole}$) of hole structure is around 408 nm, there will be a strong spectroscopic overlap between the $Q_{hole}$ and $D_{dark}$ (Compared to $Q_{hole}$, $D_{dark}$ is a bright mode), and result a Fano splitting at the place of the energy superposition. This splitting is confirmed by the extinction data of the symmetry broken nanodisk as Figure 5d, where the Fano splitting takes place exactly at the same place of the $Q_{hole}$ overlaps the $D_{dark}$ modes at 408 nm, see the pink dashed line in Figure 5. The asymmetric peak with a steeper slope toward the red than toward the blue presents the characteristics of the Fano resonance.

In conclusion, we theoretically investigated the plasmonic Fano resonance of a single symmetry broken Ag nanodisk with a normal incident laser. By increasing the open angle of the nanodisk, the Fano resonance was first enhanced and then weakened, which was controlled by the intensity of the generated quadrupolar mode of the open structure. The Fano splitting was also observed when changed the nanodisk radius from 40 to100 nm. It demonstrated that with a larger radius, the overlap between the quadrupole and bright mode was strengthened, and thus generated a strong Fano resonance. A semi-analytical method was developed to calculate the plasmon hybridization process of the dipolar modes of the structure, and also confirmed their degenerate states. The plasmon hybridization diagram presented the spectroscopic overlap between the broad dipolar mode and narrow quadrupolar mode, and the generation process for the quadru Fano resonance. The suggested symmetry broken nanodisk and its analysis method provided a useful path for the Fano line shape investigation inside of a single configuration, and can be served as plasmonic Fano-resonant sensors for the detection of nanostructure integration and deformation in the future.